# Interfacial Spin-Orbit Torques and Magnetic Anisotropy in WSe$_2$/Permalloy Bilayers

Jan Hidding[1], Sytze H. Tirion[1], Jamo Momand[1], Alexey Kaverzin[1], Maxim Mostovoy[1], Bart J. van Wees[1], Bart J. Kooi[1], and Marcos H. D. Guimarães[1]

[1] Zernike Institute of Advance Materials, University of Groningen, Groningen, The Netherlands
E-mail: jan.hidding@rug.nl and m.h.guimaraes@rug.nl

## Abstract

Transition metal dichalcogenides (TMDs) are promising materials for efficient generation of current-induced spin-orbit torques on an adjacent ferromagnetic layer. Numerous effects, both interfacial and bulk, have been put forward to explain the different torques previously observed. Thus far, however, there is no clear consensus on the microscopic origin underlying the spin-orbit torques observed in these TMD/ferromagnet bilayers. To shine light on the microscopic mechanisms at play, here we perform thickness dependent spin-orbit torque measurements on the semiconducting WSe$_2$/permalloy bilayer with various WSe$_2$ layer thickness, down to the monolayer limit. We observe a large out-of-plane field-like torque with spin-torque conductivities up to $1 \times 10^4 (\hbar/2e)(\Omega m)^{-1}$. For some devices, we also observe a smaller in-plane antidamping-like torque, with spin-torque conductivities up to $4 \times 10^3 (\hbar/2e)(\Omega m)^{-1}$, comparable to other TMD-based systems. Both torques show no clear dependence on the WSe$_2$ thickness, as expected for a Rashba system. Unexpectedly, we observe a strong in-plane magnetic anisotropy – up to about $6.6 \times 10^4$ erg/cm$^3$ – induced in permalloy by the underlying hexagonal WSe$_2$ crystal. Using scanning transmission electron microscopy, we confirm that the easy axis of the magnetic anisotropy is aligned to the armchair direction of the WSe$_2$. Our results indicate a strong interplay between the ferromagnet and TMD, and unveil the nature of the spin-orbit torques in TMD-based devices. These findings open new avenues for possible methods for optimizing the torques and the interaction with interfaced magnets, important for future non-volatile magnetic devices for data processing and storage.

# 1. Introduction

The electrical manipulation of magnetic layers is extremely appealing for future non-volatile and energy-efficient data processing and memory devices [1] [2] [3]. One of the most promising approaches to accomplish this is the use of spin-orbit torques (SOTs), where an electric current through a high spin-orbit material can apply a torque on the magnetization of an interfaced ferromagnetic layer [2]. One of the key components of materials showing large SOTs is the presence of a high spin-orbit coupling. For this reason, heavy-metal layers such as Pt [4] [5], W [6] [7], and Ta [8], have been used to generate efficient SOTs. These systems were shown to be capable of switching the direction of out-of-plane magnetic layers with relatively small current densities ($5 \times 10^5$ A/cm$^2$) [9]. For this reason, heavy-metal-based SOT devices have been in the spotlight for future magnetic random-access memory devices [2].

The application of heavy-metal layers for SOT devices has many advantages, such as the easy integration with CMOS-compatible processes [10], but it suffers from a relatively low SOT efficiency. This is partially due to their relatively weak spin Hall effect – a few 10s of percent – combined with the fact that their spin Hall-generated torques do not possess the optimum symmetry for deterministic magnetization switching of magnetic layers with perpendicular magnetic anisotropy, such as the ones used in high-density memory recording. This has pushed researchers to explore more exotic material systems, such as topological insulators and two-dimensional (2D) materials, for SOT generation. Topological insulators were shown to generate very large SOTs and magnetic switching current densities orders of magnitude lower than conventional heavy-metal devices [11] [12] [13] [14] [15] [16] [17]. However, a large portion of the current still flows through the bulk of the material and does not profit from the high spin-orbit coupling at the (topological) surface states, which reduces the SOT efficiency.

The large family of van der Waals materials, such as transition metal dichalcogenides (TMDs), have shown to be a promising material platform for the study of SOTs [18]. Due to their versatile properties, where similar materials can show drastically different values of, e.g., conductivity or spin-orbit coupling, these systems can be used to pinpoint key ingredients for effective SOT generation. The study of SOTs in TMDs with low crystal symmetries illustrates this well [19] [20] [21] [22] [23]. There, researchers identified the presence of SOTs with unusual symmetries, not allowed in conventional systems and made possible by the low symmetry of the TMD layers. The particular case of semi-metallic WTe$_2$ is very attractive since it showed a large out-of-plane antidamping-like torque [19] [20], ideal for the switching of layers with perpendicular magnetic anisotropy, in addition to large SOT efficiencies and very small critical current densities for magnetization switching [24]. The antiferromagnetic insulator $NiPS_3$ also demonstrated very large SOT efficiencies, surpassing conventional Pt/ferromagnet systems at low temperatures [25].

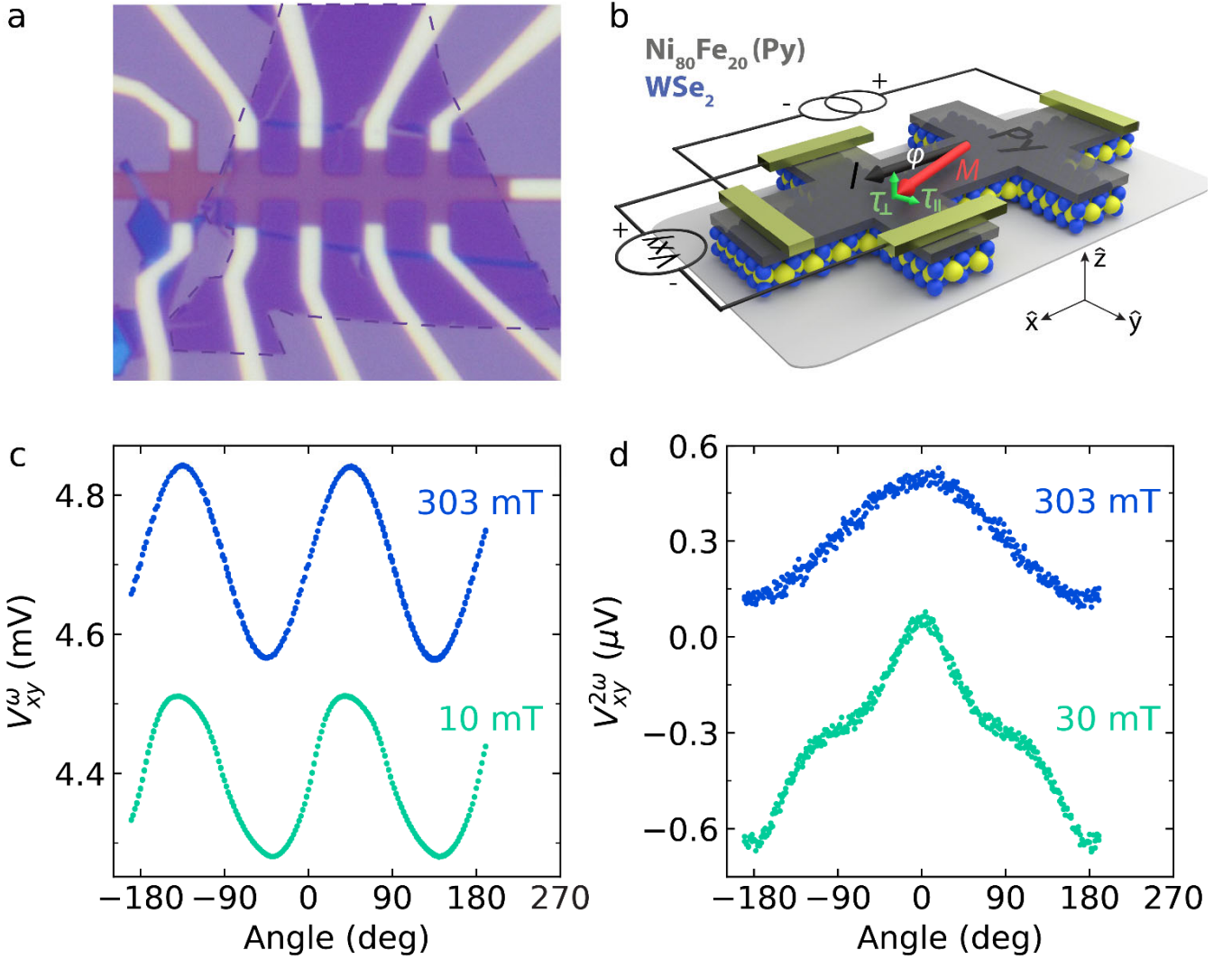

**Figure 1. (a)** Optical micrograph of a typical device ($D_2^B$) before etching the device into the Hall bar geometry. The WSe$_2$ flake is outlined by the dashed line. **(b)** Schematic of the WSe$_2$/Py device geometry and harmonic Hall measurement configuration. A current is driven along the $\hat{x}$-direction and the external magnetic field is rotated in-plane, while measuring the transverse first and second harmonic Hall voltages ($V_{xy}^{\omega/2\omega}$) along the $\hat{y}$-direction. Figure **(c)** and **(d)** show the first and second harmonic Hall voltages versus in-plane rotation of the magnetic field, respectively, at low and high external magnetic fields. Note that the signals were offset for clarity.

Semiconducting TMDs, such as hexagonal $MoS_2$ and $WSe_2$, have also attracted significant attention. These materials have shown more modest SOT efficiencies [26] [27] [28] [29] but are more attractive for industrial integration due to their air stability and the more developed methods for wafer-scale growth on CMOS-compatible substrates [30]. Even though SOTs in semiconducting TMD/ferromagnet systems have been studied before, a clear consensus on the torque symmetries and mechanisms is still lacking [26] [31] [32] [33]. Moreover, all studies performed previously used chemical vapor deposition (CVD) grown layers, which, despite providing large scale films, suffer from a lower crystalline and electronic quality compared to their mechanically exfoliated counterparts.

Here, we report current induced SOTs in high-quality single crystal $WSe_2$ interfaced with a ferromagnet, $Ni_{80}Fe_{20}$ (permalloy – Py), for multiple $WSe_2$ thicknesses, down to the monolayer limit. We observe a large out-of-plane field-like torque ($\tau_{FL}$), and, for some of our devices, a smaller in-plane antidamping-like torque ($\tau_{DL}$) with no clear dependence on the $WSe_2$ thickness for both $\tau_{FL}$ and $\tau_{DL}$. Our results are consistent with SOTs arising from an interfacial (Rashba) spin-orbit coupling. Furthermore, we observe a large magnetic anisotropy induced on the Py layer for all our devices. Two devices in particular, possessing the largest anisotropy values, allow us to identify that the magnetic anisotropy induced in the Py layer closely follows the armchair crystallographic axis of the $WSe_2$ crystal. Our study shines light on the fundamentals of SOTs in TMD/ferromagnet bilayers, making it possible to narrow down on specific microscopic mechanisms. Moreover, our observation of a large magnetic anisotropy in Py following the crystallographic axis of $WSe_2$ should further enhance the understanding of the interaction between these two materials, essential for optimizing future TMD-based spintronic devices.

## 2. Experimental Methods

*2.1 Device fabrication*

Our samples are fabricated by mechanically exfoliating a bulk $WSe_2$ crystal (HQ Graphene) on $Si/SiO_2$ [34]. Thin $WSe_2$ flakes are selected using optical microscopy and their thickness determined by optical contrast [35] and atomic-force microscopy. Monolayer flakes are further confirmed by their intense photoluminescence [36]. Only flakes with a low RMS roughness (<400 pm) and with no steps are selected to avoid artifacts in our measurements due to the roughness of the ferromagnetic layer. For this study, final devices were fabricated using WSe$_2$ flakes with a thickness ranging from monolayer to 4 layers. Subsequently, a separately prepared mask with a Hall bar opening, made on poly (methyl methacrylate) (PMMA), is dry-transferred on top the WSe$_2$ flakes to ensure a pristine interface between the WSe$_2$ flake and a 6 nm thick Py layer which is deposited by electron-beam evaporation. The Py is capped with a 17 nm thick Al$_2$O$_3$ layer to protect it from oxidation. Next, Ti/Au (5/55nm) contacts are fabricated using standard lithography and thin-film evaporation techniques. An Al$_2$O$_3$ wet etch with tetramethylammonium hydroxide is performed just before metal deposition. Finally, to create a well-defined device geometry, the WSe$_2$/Py bilayers are patterned in a Hall bar geometry using CF$_4$/O$_2$ (9.5/0.5 sccm) reactive plasma etching (30 W RF, 5 W ICP), where the Al$_2$O$_3$ layer serves as a hard mask. For most devices, the channel of our Hall bars, which establish the current direction, is defined along the edge of the flake. This is done to ensure that the current direction is applied at a nearly constant crystallographic direction. An optical image of a device before the last etching step is shown in Figure 1a.

*2.2 Electrical Measurements*

The SOTs in our devices are characterized at room temperature using conventional low-frequency harmonic Hall measurements [37] [38] [39] with currents ranging from $I_0$ = 400 to 600 μA and frequencies below 200 Hz. With this technique, a constant magnetic field μ$_0$H (10 to 300 mT) is applied in the sample plane and the sample is rotated, so the field makes an angle $\phi$ with respect to the current. Meanwhile, the first and second harmonic Hall voltages, $V_\omega^{xy}$ and $V_{2\omega}^{xy}$, respectively, are measured, as shown in Figure 1b. For a small magnetic anisotropy compared to μ$_0$H, the magnetization angle is $\phi_M \approx \phi$. The first-harmonic Hall resistance ($R_{xy}^\omega = V_\omega^{xy}/I_0$) is given by:

$$R_{xy}^\omega = R_{\text{PHE}} \sin^2(\theta) \sin(2\phi) + R_{\text{AHE}} \cos(\theta) \tag{1}$$

where $\theta$ is the magnetic field's polar angle ($\theta = 90°$ for in-plane magnetic fields), and $R_{\text{PHE(AHE)}}$ is the strength of the planar (anomalous) Hall effect resistance.

In the presence of out-of-plane field-like and in-plane damping-like SOTs ($\tau_{FL}$ and $\tau_{DL}$) and an anomalous Nernst effect voltage ($V_{ANE}$), a second Harmonic Hall voltage is generated and can be described by [4][38]:

$$V_{xy}^{2\omega}(\phi) = A\cos(2\phi)\cos(\phi) + B\cos(\phi) \tag{2}$$

where the A- and B-component are given by:

$$A = \frac{R_{\text{PHE}} I_0\ \tau_{FL}/\gamma}{H} \tag{3}$$

$$B = \frac{R_{\text{AHE}} I_0\ \tau_{DL}/\gamma}{H + H_K} + V_{ANE} \tag{4}$$

with $\gamma$ the gyromagnetic ratio, $H$ is the applied magnetic field and $H_K$ the out-of-plane anisotropy field. Due to the hexagonal symmetry of $WSe_2$, in the absence of strain, only torques with conventional symmetries are allowed [21][40]. Therefore, we expect no unconventional spin-orbit torques related to the crystal structure in our devices, which agrees with our experimental results. The SOT terms are assumed to have the conventional symmetry properties with respect to the magnetization direction ($\hat{m}$), i.e $\tau_{FL} \propto \hat{m} \times \hat{y}$ and $\tau_{DL} \propto \hat{m} \times (\hat{y} \times \hat{m})$, where $\hat{y}$ is the direction perpendicular to the current (see Figure 1b).

*2.3 Scanning Transmission Electron Microscopy*

We prepared cross-sectional specimen with a Helios G4 CX focused ion-beam (FIB) at 30 kV from Thermo Fisher Scientific, either parallel or perpendicular to the device current direction, using gradually decreasing acceleration voltages of 5 kV and 2 kV for the final polishing. Transmission electron microscopy (TEM) analyses were performed with a double aberration corrected Themis Z from Thermo Fisher Scientific, operated at 300 kV. High-angle annular dark-field scanning TEM (HAADF-STEM) images were recorded with probe currents of about 50 pA, convergence semi-angle 21 mrad or 30 mrad and HAADF collection angles 61–200 mrad.

## 3. Results

We performed harmonic Hall measurements for six $WSe_2$/Py devices with various $WSe_2$ thicknesses. For convenience, we will refer to the devices as $D_1$, $D_2^{A/B/C}$ and $D_4^{A/B}$ for the remainder of the text, where the subscript denotes the number of $WSe_2$ layers and the superscript the device label. We observe a large out-of-plane field-like torque in all but one of our $WSe_2$/Py devices with some of them showing an additional small, but measurable in-plane antidamping torque. The device that did not show a measurable SOT ($D_2^C$), however, shows unprecedently large magnetic anisotropy. As discussed in the last section, the large magnetic anisotropy counteracts the SOTs, making it very challenging to properly quantify them by our measurement technique. First, however, we will discuss the harmonic Hall measurements of the spin-orbit torques in the other five devices.

*3.1 Interfacial SOTs*

Figure 2a shows the second Harmonic Hall voltage for a two-layer (~1.4 nm thick) WSe$_2$/Py device ($D_2^A$) as a function of $\phi$ for various magnetic field strengths. The data are fitted to extract the *A* and *B* amplitudes (Eq. 2), which are then plotted as a function of the magnetic field (Fig. 2b). The presence of SOTs is revealed by the $1/H$ behavior, while the anomalous Nernst effect can be differentiated by an offset in *B*. At low fields the assumption that $H \gg H_K$, no longer holds, resulting in a worse fit and thus larger error bars. Using Eq. 3 and Eq. 4, the field-like ($\tau_{FL}$) and antidamping-like ($\tau_{DL}$) torques and are extracted, respectively.

To better quantify and compare the SOTs in our devices, we express them in terms of their spin-torque conductivities, commonly used as figure-of-merit in literature [2]. The spin-torque conductivity is defined as the angular momentum absorbed by the ferromagnet per second per applied electric field per interface area. Due to the independence of the spin-torque conductivities with respect to device geometries and resistances, it gives us a meaningful value which allows us to compare our various devices among each other as well as with values reported in literature. The spin-torque conductivities for $\tau_{FL(DL)}$ are calculated by [23][41]:

$$\sigma_{FL(DL)} = \frac{2e}{\hbar} M_s t_{Py} w \frac{\tau_{FL(DL)}/\gamma}{R_{sq} I_0} \qquad (5)$$

where $e$ and $\hbar$ are the electron charge and the reduced Planck's constant, respectively, $M_s$ is the saturation magnetization, $t_{Py}$ = 6 nm is the Py thickness, $w$ is the device width, and $R_{sq}$ is the square resistance of the WSe$_2$/Py stack. The parameters for all these devices are summarized in the supplementary information.

The spin torque conductivities for all devices versus layer thickness are shown in Figure 2c. We observe a $\sigma_{FL}$ ranging from $3.6 \pm 0.1 \times 10^3$ to $12.1 \pm 0.1 \times 10^3 \frac{\hbar}{2e} (\Omega.m)^{-1}$ and $\sigma_{DL}$ ranging from $-0.3 \pm 0.1 \times 10^3$ to $3.9 \pm 0.2 \times 10^3 \frac{\hbar}{2e} (\Omega.m)^{-1}$ with no clear correlation with the WSe$_2$ layer thickness. Our results are consistent with previous reports on CVD grown monolayer WSe$_2$/ferromagnet devices and comparable to the spin-torque conductivities for damping-like torques observed in other 2D materials [42]. Nevertheless, to the best of our knowledge, we report both the highest field-like and damping-like spin-torque conductivities found for semiconducting TMDs. This indicates that our devices possess a highly transparent interface and a large interaction between the TMD and the ferromagnet.

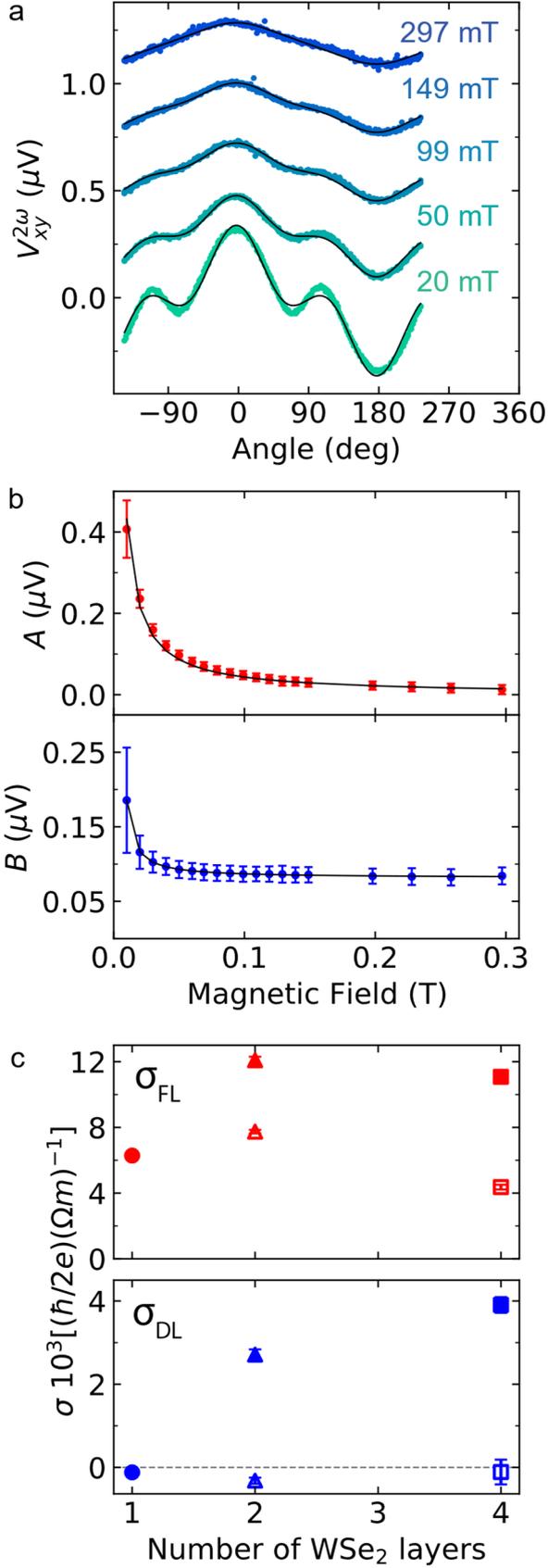

In contrast to Ref. [29], we do not observe a decreasing $\tau_{DL}$ with increasing WSe$_2$ layer thickness for our devices. Rather, we observe a seemingly increasing $\sigma_{DL}$ with increasing thickness. We note that from our five devices, only two show a sizeable $\sigma_{DL}$. Additionally, we observe a much stronger field-like torque when compared to the damping-like torque, by a factor of 6 or higher. This is consistent with some works on CVD-based WSe2 devices [27], but contrary to others [28][29]. The key differences of our process compared to the previous reports are the higher quality of $WSe_2$ crystals with single crystallographic domains obtained by mechanical exfoliation and a milder deposition of the ferromagnetic layer. Our devices show pristine interfaces between $WSe_2$ and Py, with no observable intermixing as confirmed by HAADF-STEM cross-sectional imaging – see below. Moreover, increase scattering at the interface has been predicted to give an increase damping-like torque, which could explain the large DL torque observed previously [28][29].

**Figure 2. (a)** Second harmonic Hall measurements at various magnetic fields for one WSe$_2$/Py device $D_2^A$. The colored circle represent the data while the black line are fits using Eq. (2). The measurements for different fields have been offset for clarity. **(b)** The A- and B-component extracted from the fit of (a) for various magnetic field strengths. The black line corresponds to the fits using Eq. (3) and Eq. (4). The error bars are obtained from the standard deviation from the fit. **(c)** Spin-torque conductivities calculated using Eq. (5) for devices with various WSe$_2$ thickness. The different symbols correspond to the different devices.

Therefore, we expect a cleaner interface quality to be the main reason for these variations.

As Py is known to show torques even in the absence of other spin-orbit materials [41], we fabricated control Py/$Al_2O_3$ samples to exclude that the strong field-like torque is generated solely in the Py/$Al_2O_3$ layer. Similar to previous reports [29][41], we observe a non-zero field-like torque, with an average spin-torque conductivity of $\sigma_{FL} = (-2.5 \pm 0.1) \times 10^3 \frac{\hbar}{2e}(\Omega m)^{-1}$. Note that we find a negative spin-torque conductivity, showing that these torques have the opposite direction as the ones we measure in our $WSe_2$ devices. The sign difference indicates that the field-like torque in the $WSe_2$/Py devices reported here are most likely an underestimation of the torques produced at the $WSe_2$/Py interface as they compete with opposite torques produced at the Py/$Al_2O_3$ interface. No significant damping-like torque was observed in these Py/$Al_2O_3$ samples (see supplementary information for details).

The absence of a thickness dependence in our devices for the field-like torque indicates that the torque does not originate from current-induced Oersted fields, for which an increasing torque is expected with increasing layer thickness. This is also in agreement with most of the current flowing through the Py layer due to its much higher conductivity when compared to $WSe_2$. A simple estimation suggests that our FL torques could only be explained by an unreasonable large conductivity for $WSe_2$, of $\sigma_{WSe_2} \sim 10^6 \, (\Omega m)^{-1}$, about 5 orders of magnitude higher than literature values [43]. Moreover, we point out that the sign of the FL torque we obtain is opposite to the one expected from a current flowing through the $WSe_2$ layer. We confirm the sign of the Oersted torques by control Pt/Py devices. We note that we observe no variation of the SOTs with gate voltages ranging up to $\pm 60$ V (equivalent to electric fields of $\pm 2.1$ MV/cm) (see supplementary information), similar to previous reports [29], which could be due to a large Schottky barrier [44] or Fermi level pinning at the metal-semiconductor interface [45].

The independence of SOT strength with WSe$_2$ thickness and the larger $\tau_{FL}$ in comparison to $\tau_{DL}$ are consistent with interfacial SOTs [46] [47] [48]. In systems with an interfacial Rashba-type spin-orbit coupling, a pure out-of-plane field-like torque is expected. However, it has been theoretically shown that an in-plane damping-like torque can arise in the presence of electron scattering [31] [33] [46] [47] [48]. Therefore, our data indicate that the SOTs in our devices arise from a Rashba-type spin-orbit coupling at the interface, with some devices showing a stronger scattering, giving rise to $\tau_{DL}$. The variation in spin torque conductivities between the devices is ascribed to a difference in WSe$_2$/Py interface quality. Variations in interface quality can affect the spin transparency of the interface, resulting in differences in the torque strength. Furthermore, pristine interfaces are expected to show stronger Rashba-effects. This is in line with our observation that the devices showing a measurable $\tau_{DL}$ also possess the highest $\tau_{FL}$. Furthermore, the importance of the interface quality is highlighted by the fact that we observe no measurable SOTs when using standard lithography techniques to fabricate similar devices, where no particular care to maintain a pristine interface was taken (see supplementary information for details).

Previous reports show that the second-harmonic signal might be falsely attributed to a damping-like torque $\tau_{DL}$ in cases of significant unidirectional magnetoresistance (UMR) arising from electron-magnon scattering [42]. To verify this, we measured the second-harmonic longitudinal voltage and find a similar second-harmonic signal, with a $\sin(\phi)$ behavior and $1/H$ dependence. This is expected since the $V_{xx}^{2\omega}$ is phase shifted relative to $V_{xy}^{2\omega}$ by 90°. Therefore, a contribution from UMR to our second-harmonic signal cannot be ruled out. However, assuming similar interface properties for our devices, as hinted by the similar values for the FL torque, we expect similar contributions from UMR to our signals used for evaluation of the DL torque. Therefore, we believe that UMR alone cannot explain the varying signal we attribute to DL torque in our devices.

*3.2 Magnetic Anisotropy*

One of the most striking differences between our devices, consisting of exfoliated WSe$_2$ crystals, and previous studies based on CVD-grown films is the presence of a strong magnetic anisotropy induced in the Py film. The first harmonic Hall voltages for our devices are expected to follow a $\sin(2\phi)$ behavior due to the planar Hall effect, Eq. (1). However, for all devices, at low external magnetic fields ($< 20$ mT), we observe clear deviations from the planar Hall effect, Figure 1c (device $D_2^B$). For device $D_1$, and in particular device $D_2^C$, we observe very strong deviations from the expected $\sin(2\phi)$ behavior even for external magnetic fields up to 100 mT. This demonstrates a very strong induced magnetic anisotropy in the Py layer, as shown in Figure 3a for device $D_2^C$, causing the magnetization angle of Py, $\phi_M$, to slightly deviate from the applied magnetic field angle, $\phi$.

To study the magnetic anisotropy in more detail, we modify Eq. (1) to account for an in-plane uniaxial anisotropy, with strength $H_A$ much smaller than the applied magnetic field and an easy-axis angle $\phi_E$ with respect to the current [20]:

$$R_\omega = R_{PHE} \sin(2\phi_M), \tag{6}$$

with

$$\phi_M = \phi - \frac{H_A}{2H} \sin[2(\phi - \phi_E)]. \tag{7}$$

For all our devices, apart from $D_1$ and $D_2^C$, we observe $H_A \approx 0.01$ to $0.16$ T, and a $\phi_E \approx 0°$ or $\pm 30°$, hinting towards a relation between the magnetic anisotropy direction in the Py and the hexagonal crystal structure of the underlying WSe$_2$. The values we find for $H_A$ for these devices are higher by factors of 2 to 10 than for those reported in similar systems [19] [20] [22] [23]. Our results for all devices are summarized in the supplementary information.

For devices $D_1$ and $D_2^C$, showing much stronger anisotropy, we find stronger deviations from the fits using equations 6 and 7. Due to the stronger anisotropy, the approximation $H_A \ll H$ taken above is no longer valid. Therefore, we use a simple macrospin model to fit the data. First, we find the magnetization angle $\phi_M$ at an applied magnetic field angle $\phi_H$, by minimizing the magnetic energy:

$$E = \left(\frac{K_2}{2}\right) \cos(2\phi_M - 2\phi_E) - h \cos(\phi_M - \phi_H) \tag{8}$$

where $K_2$ is the 2-fold anisotropy constant, and $h$ is the Zeeman energy by the applied magnetic field at an angle $\phi_H$. We find that our data agrees with this simple theoretical model with $K_2 = 2 \times 10^4$ erg/cm$^3$ and $6.6 \times 10^4$ erg/cm$^3$ for devices $D_1$ and $D_2^C$, respectively. For devices $D_1$ and $D_2^C$ we find $\phi_E = -60°$ ($D_1$) and $\phi_E = -28°$ ($D_2^C$), with respect to the current direction. This suggests that the induced magnetic anisotropy is indeed related to the hexagonal crystal structure of the WSe$_2$.

To confirm the crystallographic direction and the interface quality of our devices, we performed HAADF-STEM cross-sectional imaging in two devices and two additional WSe$_2$ flakes. Figure 3c shows a cross-sectional HAADF-STEM image of device $D_2^C$. The two layers of the WSe$_2$ are visible with atomic resolution and the STEM image reveals the randomly oriented polycrystalline structure of the Py layer on top. The STEM image shows an atomically sharp interface indicating a clean interface for our fabricated devices. In addition, little to no intermixing is observed demonstrating that the WSe$_2$ layer experiences little to no damage upon evaporation of the Py layer and that the crystalline orientation of the layer remains uniform.

For the crystallographic direction, we find that the edge of all WSe$_2$ flakes studied by STEM follow a zigzag direction. For device $D_2^C$, the Hall bar channel is aligned with the edge of the WSe$_2$ flake, so that the current flows along the zigzag direction (schematically indicated in Figure 3d). As we found $\phi_E = -28°$ for device $D_2^C$ following our analysis, we conclude that the (uniaxial) magnetic anisotropy observed in the Py layer lies along the armchair direction of the WSe$_2$. For device $D_1$, the Hall bar channel is aligned 30° away from the edge of the WSe$_2$ flake, in which case the current flows along the armchair direction. Correspondingly, we find that $\phi_E = -60°$, which again shows that the (uniaxial) magnetic anisotropy is aligned with the armchair direction of the WSe$_2$.

The correspondence between the magnetic anisotropy direction in Py and the crystallographic directions of the WSe$_2$ indicates a strong interaction between the two materials. Similar results have been found in low-symmetry TMDs, such as WTe$_2$ [19][20], MoTe$_2$ [23], and TaTe$_2$ [22], with values around $H_A \sim 10$ mT, one order of magnitude lower compared to values observed in our devices (supplementary information Table 1). However, systems with higher (hexagonal) symmetries, such as NbSe$_2$ [21] and NiPS$_3$ [25], did not show such effects, even though in the case of NbSe$_2$, the symmetry might have been reduced by strain. We point out that the polycrystallinity of CVD-grown crystals would not allow for such observation, which is supported by the lack of magnetic anisotropy in previous studies [27][28][29]. Moreover, such large values of the magnetic anisotropy shown by devices $D_1$ and $D_2^C$ are unprecedented in TMD-based devices. We do not fully understand the differences in anisotropy strength between devices $D_1$ and $D_2^C$ and the other devices since all device fabrication steps were identical. Nevertheless, due to the strong dependence of the magnetic anisotropy on the interface quality and the fact that particular care was taken in maintaining a clean interface during fabrication, we have arguments to believe that our devices have more pristine interfaces, resulting in a stronger interaction between Py and WSe$_2$.

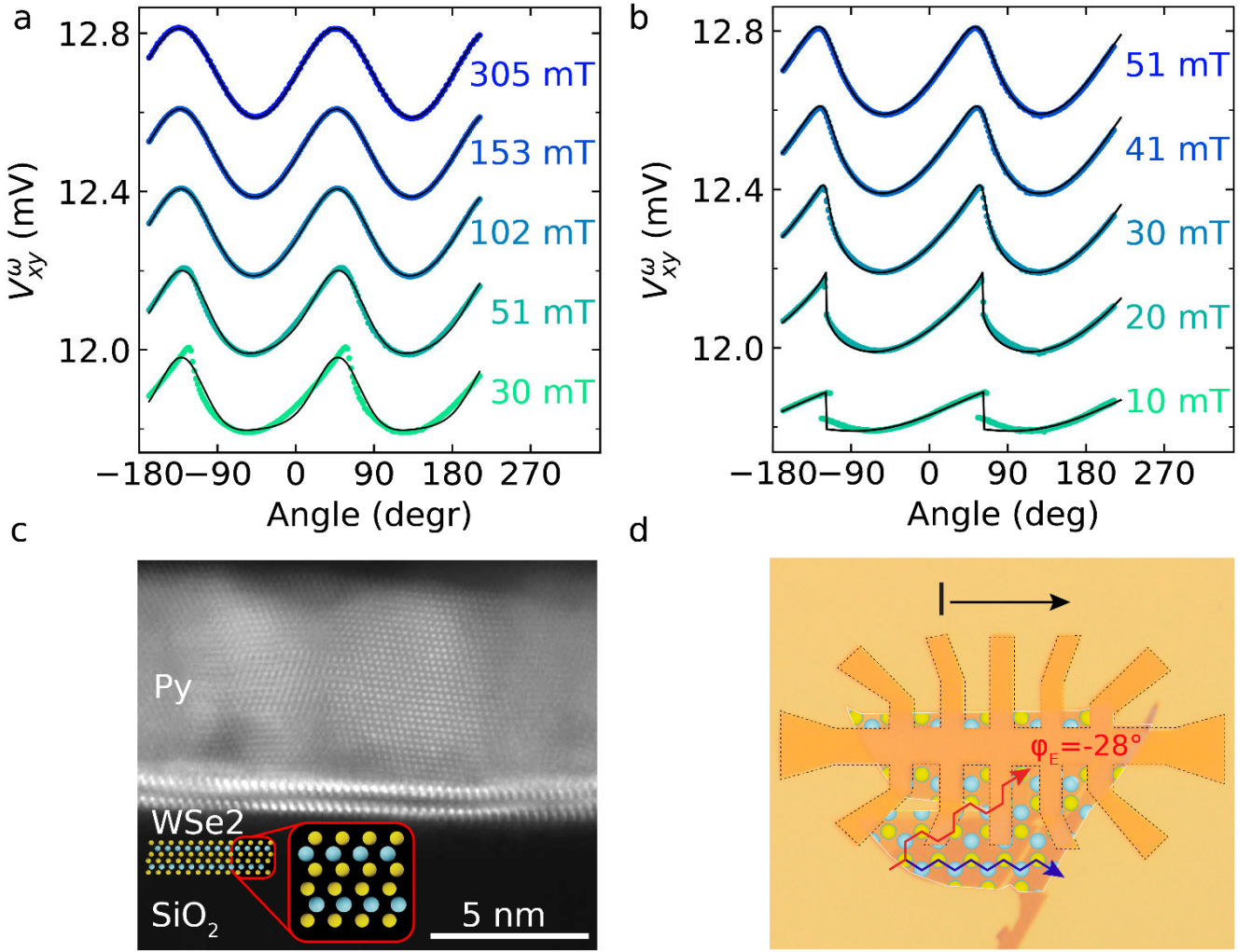

**Figure 3. (a)** First harmonic Hall voltage for one of the WSe$_2$/Py devices (D$_2^c$), shown in (d). The black line corresponds to the fit using Eq. (6) and (7) and colored circles to the experimental data. Clear deviations from the planar Hall effect are apparent, indicating an in-plane magnetic anisotropy in the Py layer. **(b)** First harmonic Hall voltage for the same device as in (a) for low magnetic fields, fitted with the model discussed in the main text (black line). Note that the plots in (a) and (b) have been given an offset for clarity. **(c)** Cross sectional HAADF-STEM image of the device shown in (d). The two layers of WSe$_2$ are clearly identified with atomic resolution. The cross section is made perpendicular to the current direction in between the legs of the Hall bar. The crystallographic orientation of the WSe$_2$ layer is apparent from the similarities between the schematic inset, showing the cross section of the armchair direction, and the STEM image, which indicates that the current direction is along the zigzag direction, as shown in (d). **(d)** Micrograph of the same device in (a) with the hexagonal crystal structure of WSe$_2$ overlayed on the WSe$_2$ crystal (outlined in white). The blue and red arrows indicate the zigzag and armchair direction, respectively. Notice that the channel of the Hall bar (black dashed line) is aligned with one of the cleaving directions of the WSe$_2$ crystal and the current direction is therefore along the flake edge for device D$_2^c$.

## 4. Conclusions

In conclusion, we report large field-like torques in WSe$_2$/permalloy bilayers, with an additional antidamping-like torque observed in a few devices and no clear dependence on WSe$_2$ layer thickness. The appearance of a weaker damping-like torque in these systems confirms the prediction of recent theoretical work on similar interfacial Rashba systems including scattering and accentuates the importance of the heavy metal/ferromagnet interface quality for tailoring towards highly efficient torques. Furthermore, we observe an induced uniaxial magnetic anisotropy in all our devices, with two devices in particular showing very strong anisotropy, aligned with the armchair direction of the WSe$_2$. Although smaller magnetic anisotropy values were observed in low-symmetry TMD-based devices, a microscopic understanding of the mechanisms involved is still lacking. The higher crystal symmetry of WSe$_2$ combined with the larger anisotropy values we observe, are expected to help researchers to develop a more detailed theoretical description of this phenomenon and, eventually, a better understanding of all the effects regarding crystal symmetry involved for spin-orbit torque devices. The knowledge on the microscopic mechanisms at play, for both SOTs and induced magnetic anisotropy, should shine light on the factors required for the development of more efficient devices for data processing and storage.

## Acknowledgements

We would like to thank Ismail El Baggari for his valuable suggestions on the sample preparation for the cross-sectional HAADF-STEM images. We thank J. G. Holstein, H. Adema, H. de Vries, A. Joshua for their technical support. This work was supported by the Dutch Research Council (NWO – STU.019.014), the Zernike Institute for Advanced Materials, the European Union Horizon 2020 research and innovation program under grant agreement No. 881603 (Graphene Flagship) and the Spinoza Prize awarded to BJ van Wees by NWO. The device fabrication was performed using NanoLabNL facilities.

## References

[1] B. Dieny *et al.*, "Opportunities and challenges for spintronics in the microelectronics industry," *Nat. Electron.*, vol. 3, no. August, 2020, doi: 10.1038/s41928-020-0461-5.

[2] A. Manchon *et al.*, "Current-induced spin-orbit torques in ferromagnetic and antiferromagnetic systems,"

*Rev. Mod. Phys.*, vol. 91, no. 3, pp. 1–72, Sep. 2018, doi: 10.1103/RevModPhys.91.035004.

[3] A. Manchon, H. C. Koo, J. Nitta, S. M. Frolov, and R. A. Duine, "New perspectives for Rashba spin-orbit coupling," *Nat. Mater.*, vol. 14, no. 9, pp. 871–882, 2015, doi: 10.1038/nmat4360.

[4] K. Garello *et al.*, "Symmetry and magnitude of spin–orbit torques in ferromagnetic heterostructures," *Nat. Nanotechnol.*, vol. 8, no. 8, pp. 587–593, Aug. 2013, doi: 10.1038/nnano.2013.145.

[5] L. Liu, T. Moriyama, D. C. Ralph, and R. A. Buhrman, "Spin-Torque Ferromagnetic Resonance Induced by the Spin Hall Effect," *Phys. Rev. Lett.*, vol. 106, no. 036601, pp. 1–4, 2011, doi: 10.1103/PhysRevLett.106.036601.

[6] C.-F. Pai, L. Liu, Y. Li, H. W. Tseng, D. C. Ralph, and R. A. Buhrman, "Spin transfer torque devices utilizing the giant spin Hall effect of tungsten," *Appl. Phys. Lett.*, vol. 101, no. 122404, pp. 1–4, 2012, doi: https://doi.org/10.1063/1.4753947.

[7] K. Demasius *et al.*, "Enhanced spin–orbit torques by oxygen incorporation in tungsten film," *Nat. Commun.*, vol. 7, no. 10644, pp. 1–7, 2016, doi: 10.1038/ncomms10644.

[8] L. Liu, C. Pai, Y. Li, H. W. Tseng, D. C. Ralph, and R. A. Buhrman, "Spin-Torque Switching with the Giant Spin Hall Effect of Tantalum," *Science (80-. ).*, vol. 336, no. 6081, pp. 555–559, 2012, doi: 10.1126/science.1218197.

[9] V. Ostwal, T. Shen, and J. Appenzeller, "Efficient Spin-Orbit Torque Switching of the Semiconducting Van Der Waals Ferromagnet $Cr_2Ge_2Te_6$," *Adv. Mater.*, vol. 32, no. 7, pp. 1–7, 2020, doi: 10.1002/adma.201906021.

[10] K. Garello *et al.*, "Manufacturable 300mm platform solution for Field-Free Switching SOT-MRAM," *Symp. VLSI Circuits*, pp. T194–T195, 2019, doi: 10.23919/VLSIC.2019.8778100.

[11] A. R. Mellnik *et al.*, "Spin-transfer torque generated by a topological insulator," *Nature*, vol. 511, no. 7510, pp. 449–451, Jul. 2014, doi: 10.1038/nature13534.

[12] Y. Wang *et al.*, "Topological Surface States Originated Spin-Orbit Torques in $Bi_2Se_3$," *Phys. Rev. Lett.*, vol. 114, no. 25, p. 257202, Jun. 2015, doi: 10.1103/PhysRevLett.114.257202.

[13] Y. Wang *et al.*, "Room temperature magnetization switching in topological insulator-ferromagnet

heterostructures by spin-orbit torques," *Nat. Commun.*, vol. 8, no. 1, p. 1364, Dec. 2017, doi: 10.1038/s41467-017-01583-4.

[14] J. Han, A. Richardella, S. A. Siddiqui, J. Finley, N. Samarth, and L. Liu, "Room-Temperature Spin-Orbit Torque Switching Induced by a Topological Insulator," *Phys. Rev. Lett.*, vol. 119, no. 077702, pp. 1–5, 2017, doi: 10.1103/PhysRevLett.119.077702.

[15] Y. Fan *et al.*, "Magnetization switching through giant spin–orbit torque in a magnetically doped topological insulator heterostructure," *Nat. Mater.*, vol. 13, no. July, pp. 699–704, 2014, doi: 10.1038/NMAT3973.

[16] N. H. D. Khang, Y. Ueda, and P. N. Hai, "A conductive topological insulator with large spin Hall effect for ultralow power spin–orbit torque switching," *Nat. Mater.*, vol. 17, no. September, pp. 808–814, 2018, doi: https://doi.org/10.1038/s41563-018-0137-y A.

[17] J. Han and L. Liu, "Topological insulators for efficient spin–orbit torques," *APL Mater.*, vol. 9, no. 6, p. 060901, 2021, doi: 10.1063/5.0048619.

[18] J. Hidding and M. H. D. Guimarães, "Spin-Orbit Torques in Transition Metal Dichalcogenide/Ferromagnet Heterostructures," *Front. Mater.*, vol. 7, no. November, 2020, doi: 10.3389/fmats.2020.594771.

[19] D. MacNeill, G. M. Stiehl, M. H. D. Guimaraes, R. A. Buhrman, J. Park, and D. C. Ralph, "Control of spin-orbit torques through crystal symmetry in WTe2/ferromagnet bilayers," *Nat. Phys.*, vol. 13, no. 3, pp. 300–305, 2017, doi: 10.1038/nphys3933.

[20] D. Macneill, G. M. Stiehl, M. H. D. Guimarães, N. D. Reynolds, R. A. Buhrman, and D. C. Ralph, "Thickness dependence of spin-orbit torques generated by WTe2," *Phys. Rev. B*, vol. 96, no. 5, pp. 1–8, 2017, doi: 10.1103/PhysRevB.96.054450.

[21] M. H. D. Guimarães *et al.*, "Spin-Orbit Torques in NbSe2/Permalloy Bilayers," *Nano Lett.*, vol. 18, no. 2, pp. 1311–1316, 2018, doi: 10.1021/acs.nanolett.7b04993.

[22] G. M. Stiehl *et al.*, "Current-induced torques with dresselhaus symmetry due to resistance anisotropy in 2D materials," *ACS Nano*, vol. 13, no. 2, pp. 2599–2605, 2019, doi: 10.1021/acsnano.8b09663.

[23] G. M. Stiehl *et al.*, "Layer-dependent spin-orbit torques generated by the centrosymmetric transition metal

dichalcogenide β-MoTe2," *Phys. Rev. B*, vol. 100, no. 18, p. 184402, Nov. 2019, doi: 10.1103/PhysRevB.100.184402.

[24] S. Shi *et al.*, "All-electric magnetization switching and Dzyaloshinskii–Moriya interaction in WTe2/ferromagnet heterostructures," *Nat. Nanotechnol.*, vol. 14, pp. 945–949, 2019, doi: 10.1038/s41565-019-0525-8.

[25] C. F. Schippers, H. J. M. Swagten, and M. H. D. Guimarães, "Large interfacial spin-orbit torques in layered antiferromagnetic insulator NiPS3/ferromagnet bilayers," *Phys. Rev. Mater.*, vol. 4, no. 084007, pp. 1–8, 2020, doi: 10.1103/PhysRevMaterials.4.084007.

[26] K. Dolui and B. K. Nikolic, "Spin-orbit-proximitized ferromagnetic metal by monolayer transition metal dichalcogenide: Atlas of spectral functions, spin textures and spin-orbit torques in Co/MoSe2, Co/WSe2 and Co/TaSe2 heterostructures," *Phys. Rev. Mater.*, vol. 4, no. 104007, pp. 1–12, 2020, [Online]. Available: http://arxiv.org/abs/2006.11335.

[27] Q. Shao *et al.*, "Strong Rashba-Edelstein Effect-Induced Spin−Orbit Torques in Monolayer Transition Metal Dichalcogenide/Ferromagnet Bilayers," *Nano Lett.*, no. 16, p. 7514−7520, 2016, doi: 10.1021/acs.nanolett.6b03300.

[28] W. Zhang *et al.*, "Research Update: Spin transfer torques in permalloy on monolayer MoS2," *APL Mater.*, vol. 4, no. 3, 2016, doi: 10.1063/1.4943076.

[29] S. Novakov, B. Jariwala, N. M. Vu, A. Kozhakhmetov, J. A. Robinson, and J. T. Heron, "Interface Transparency and Rashba Spin Torque Enhancement in WSe2Heterostructures," *ACS Appl. Mater. Interfaces*, vol. 13, p. 13, 2021, doi: 10.1021/acsami.0c19266.

[30] K. Kang *et al.*, "High-mobility three-atom-thick semiconducting films with wafer-scale homogeneity," *Nature*, vol. 520, pp. 656–660, 2015, doi: 10.1038/nature14417.

[31] F. Sousa, G. Tatara, and A. Ferreira, "Skew-scattering-induced giant antidamping spin-orbit torques: Collinear and out-of-plane Edelstein effects at two-dimensional material/ferromagnet interfaces," *Phys. Rev. Res.*, vol. 2, no. 4, pp. 1–10, 2020, doi: 10.1103/physrevresearch.2.043401.

[32] L. Zhu, D. C. Ralph, and R. A. Buhrman, "Spin-Orbit Torques in Heavy-Metal-Ferromagnet Bilayers with17

Varying Strengths of Interfacial Spin-Orbit Coupling," *Phys. Rev. Lett.*, vol. 122, no. 077201, 2019, doi: 10.1103/PhysRevLett.122.077201.

[33] K. Zollner, M. D. Petrović, K. Dolui, P. Plecháč, B. K. Nikolić, and J. Fabian, "Scattering-induced and highly tunable by gate damping-like spin-orbit torque in graphene doubly proximitized by two-dimensional magnet Cr2Ge2Te6 and monolayer WS2," *Phys. Rev. Res.*, vol. 2, no. 4, pp. 1–11, 2020, doi: 10.1103/physrevresearch.2.043057.

[34] K. S. Novoselov *et al.*, "Two-dimensional atomic crystals," *PNAS*, vol. 102, no. 30, pp. 10451–10453 PHYSICS, 2005.

[35] A. Castellanos-Gomez, N. Agraït, and G. Rubio-Bollinger, "Optical identification of atomically thin dichalcogenide crystals," *Appl. Phys. Lett.*, vol. 96, no. 213116, 2010, doi: 10.1063/1.3442495.

[36] K. F. Mak, C. Lee, J. Hone, J. Shan, and T. F. Heinz, "Atomically thin MoS2: A new direct-gap semiconductor," *Phys. Rev. Lett.*, vol. 105, no. 13, pp. 2–5, 2010, doi: 10.1103/PhysRevLett.105.136805.

[37] K. Garello *et al.*, "Symmetry and magnitude of spin-orbit torques in ferromagnetic heterostructures," *Nat. Nanotechnol.*, vol. 8, no. 8, pp. 587–593, 2013, doi: 10.1038/nnano.2013.145.

[38] M. Hayashi, J. Kim, M. Yamanouchi, and H. Ohno, "Quantitative characterization of the spin-orbit torque using harmonic Hall voltage measurements," *Phys. Rev. B - Condens. Matter Mater. Phys.*, vol. 89, no. 14, p. 144425, Apr. 2014, doi: 10.1103/PhysRevB.89.144425.

[39] M. H. Nguyen and C. F. Pai, "Spin-orbit torque characterization in a nutshell," *APL Mater.*, vol. 9, no. 3, 2021, doi: 10.1063/5.0041123.

[40] Y. Liu and Q. Shao, "Two-Dimensional Materials for Energy- E ffi cient Spin − Orbit Torque Devices," *ACS Nano*, vol. 14, pp. 9389–9407, 2020, doi: 10.1021/acsnano.0c04403.

[41] M. Nguyen, D. C. Ralph, and R. A. Buhrman, "Spin Torque Study of the Spin Hall Conductivity and Spin Diffusion Length in Platinum Thin Films with Varying Resistivity," *Phys. Rev. Lett.*, vol. 116, no. 12, p. 126601, Mar. 2016, doi: 10.1103/PhysRevLett.116.126601.

[42] Q. Shao *et al.*, "Strong Rashba-Edelstein Effect-Induced Spin-Orbit Torques in Monolayer Transition Metal Dichalcogenide/Ferromagnet Bilayers," *Nano Lett.*, vol. 16, no. 12, pp. 7514–7520, 2016, doi:

10.1021/acs.nanolett.6b03300.

[43]  M. K. Agarwal, A. R. Jani, J. D. Kshtriya, M. N. Vashi, and P. K. Garg, "Some Transport Measurements of WSe2-x Single Crystals," *Cryst. Res. Technol.*, vol. 19, no. 12, pp. 1575–1582, 1984.

[44]  R. T. Tung, "The physics and chemistry of the Schottky barrier height," *Appl. Phys. Rev.*, vol. 1, no. 1, 2014, doi: 10.1063/1.4858400.

[45]  K. Sotthewes *et al.*, "Universal Fermi-Level Pinning in Transition-Metal Dichalcogenides," *J. Phys. Chem. C*, vol. 123, no. 9, pp. 5411–5420, 2019, doi: 10.1021/acs.jpcc.8b10971.

[46]  V. P. Amin and M. D. Stiles, "Spin transport at interfaces with spin-orbit coupling: Phenomenology," *Phys. Rev. B*, vol. 94, no. 10, p. 104420, Sep. 2016, doi: 10.1103/PhysRevB.94.104420.

[47]  V. P. Amin and M. D. Stiles, "Spin transport at interfaces with spin-orbit coupling: Formalism," *Phys. Rev. B*, vol. 94, no. 10, p. 104420, Sep. 2016, doi: 10.1103/PhysRevB.94.104419.

[48]  V. P. Amin, P. M. Haney, and M. D. Stiles, "Interfacial spin–orbit torques," *J. Appl. Phys.*, vol. 128, no. 151101, 2020, doi: 10.1063/5.0024019.